\documentclass[conference]{IEEEtran}
\usepackage{graphicx}
\usepackage{float}
\usepackage{multirow}
\usepackage{diagbox}

\usepackage{lipsum}
\newcommand\blfootnote[1]{%
	\begingroup
	\renewcommand\thefootnote{}\footnote{#1}%
	\addtocounter{footnote}{-1}%
	\endgroup
}

\makeatletter
\def\footnoterule{\kern-3\p@
	\hrule \@width 2in \kern 2.6\p@} 
\makeatother

\usepackage{url}

\ifCLASSINFOpdf
\else
\fi
\begin{document}
%
\title{Quick NAT: High Performance NAT System \\on Commodity Platforms }

\author{
	\IEEEauthorblockN{Junfeng Li, Dan Li, Yukai Huang, Yang Cheng, Ruilin Ling}
	\IEEEauthorblockA{Tsinghua National Laboratory for Information Science and Technology\\
	Department of Computer Science and Technology, Tsinghua University, China\\
	hotjunfeng@163.com, \{huangyk14, cheng-y16, lrl14\}@mails.tsinghua.edu.cn, tolidan@tsinghua.edu.cn
	}
}

\maketitle

\begin{abstract}
NAT gateway is an important network system in today's IPv4 network when translating a private IPv4 address to a public address. However, traditional NAT system based on Linux Netfilter cannot achieve high network throughput to meet modern requirements such as data centers. To address this challenge, we improve the network performance of NAT system by three ways. First, we leverage DPDK to enable polling and zero-copy delivery, so as to reduce the cost of interrupt and packet copies. Second, we enable multiple CPU cores to process in parallel and use lock-free hash table to minimize the contention between CPU cores. Third, we use hash search instead of sequential search when looking up the NAT rule table. Evaluation shows that our Quick NAT system significantly improves the performance of NAT on commodity platforms.
\blfootnote{This is the author's version of the work (updated on May 9th, 2021). It is posted here only for your personal use. Not for redistribution.}
\blfootnote{\url{https://doi.org/10.1109/LANMAN.2017.7972137}}
\end{abstract}


\section{Introduction}
The scale of Internet is ever increasing. Today there are more than 1 billion hosts connected to Internet. Since IPv4 addresses are exhausted in 2011~\cite{exhausted}, the move to IPv6 seems inevitable. However, the transition from IPv4 to IPv6 requires updating not only the Internet infrastructure but also a large amount of applications, which faces many obstacles in practice. As a result, IPv4 network and IPv4 users are still the dominant in today's Internet. In IPv4 network, the key technology to deal with the address insufficiency problem is NAT (Network Address Translation). 

In the past, most NAT systems are deployed in the Linux platform leveraging the Netfilter framework \cite{netfilter}. Although it may work in small-scale networks, its performance faces significant challenge with high traffic volume. Specifically, for small-sized packets, the throughput of NAT system on commodity servers can hardly exceed 1Gbps, which leads to a big gap between the system performance and the hardware capability with 10G/100G NIC cards and multiple CPU cores. In this work we try to improve the performance of NAT system by the following approaches. First, we leverage the DPDK (Data Plane Development Kit)'s capabilities to build NAT system in the user space instead of in the Linux kernel, and thus enable polling the NIC to read packets directly into user space to eliminate the high overhead caused by packet copy and interrupt. But we also need to manipulate the packet through pointers to achieve zero-copy in the process of NAT. Second, to leverage the multi-core capability of modern commodity servers, we enable RSS (Receive-side Scaling) to let multiple cores process packets in a parallel way. But we need to minimize the sharing cost between CPU cores. Third, we find that the algorithms used in today's NAT system can also be improved. In particular, we use hash based search instead of sequential search when looking up the NAT rule table, which also considerably helps improve the performance.

Based on the improvements above, we implement a NAT system called Quick NAT. Our experiments show that Quick NAT system significantly reduces the time to search for NAT rules.

\section{System Design}
In order to improve the performance of NAT on commodity platforms, we design Quick NAT system built on DPDK \cite{DPDK}. 

\subsection{System Overview}
The architecture of Quick NAT system is shown in Figure 1. Quick NAT system utilizes DPDK's capabilities to bypass the kernel and be built in the user space. Quick NAT system is composed of four components, i.e. Connection Tracer, Rule Finder, Tuple Rewriter and IP/Port Pool. 

\begin{figure}[ht!] 
\centering
\includegraphics[width = .4\textwidth]{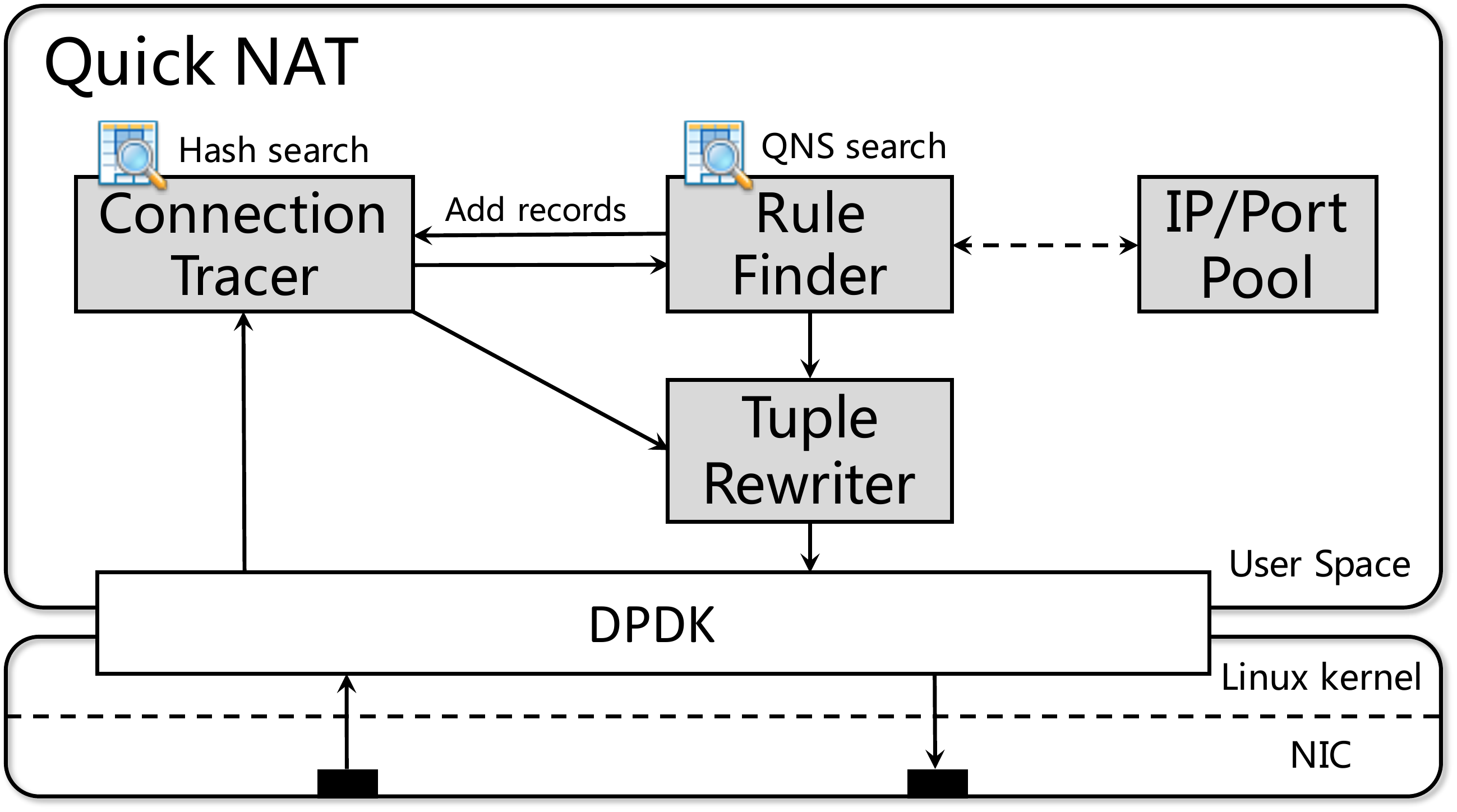}
\caption{Quick NAT system architecture}
\label{RSG}
\end{figure}

In Quick NAT system, we have made three major contributions to improve the performance of NAT.

\subsection{QNS Algorithm}
We design QNS (Quick NAT Search) Algorithm to reduce the time of NAT rule lookup. It uses hash search instead of sequential search to look up NAT rule tables with complexity of O(1).

Quick NAT system uses small rule tables (i.e. 32 DNAT rule tables and 32 SNAT rule tables) instead of one big NAT rule table. NAT rules are stored into different small rule tables according to the subnet mask and NAT-type. Moreover, each small NAT rule table has one bit as flag to indicate whether it contains rules. QNS searches different rule tables one by one in the order of decreasing subnet mask because the rule with longer subnet mask bits is preciser than that with lower mask. 


We use an example to illustrate how QNS works. As Figure 2 shows, QNS starts with the SNAT rule table with 32-bit subnet mask. QNS computes hash value on the basis of source IP and port of this packet. In this scenario, the hash value is 1288 and thus it does not find a SNAT rule with exact port and 32-bit subnet mask. And then QNS computes hash value based on source IP and zero port to search SNAT rules with wildcard port and still does not find the rule with wildcard port to match this packet. In all, QNS does not find the rule to match this packet in this sub-table.
\begin{figure}[ht!]
\centering
\includegraphics[width = .48\textwidth]{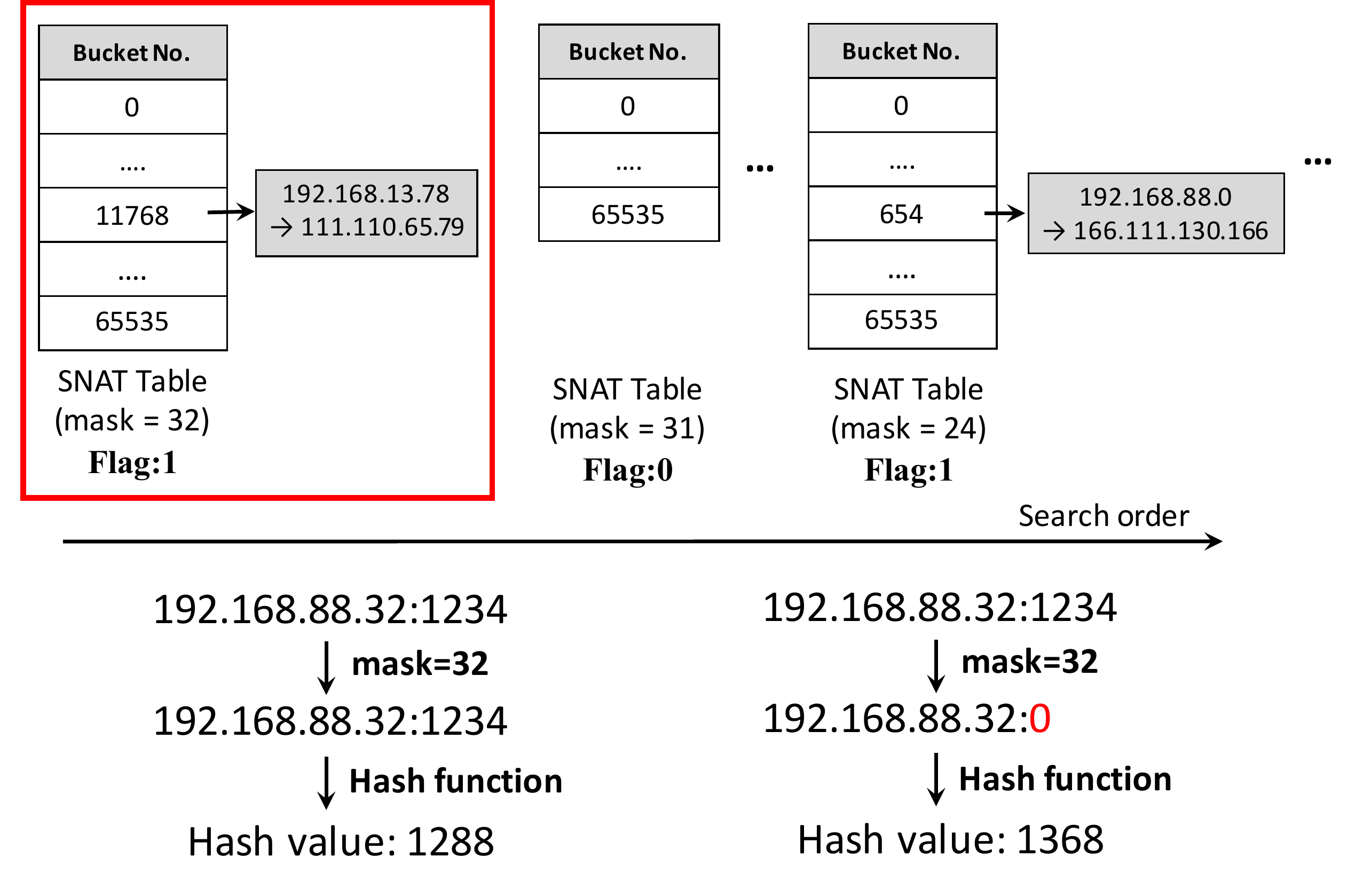}
\caption{Search for SNAT rule (step 1)}
\label{RSG}
\end{figure}

If a rule table's flag bit is zero, there is no rule in this table and QNS skips this sub-table. 

Figure 3 shows that QNS turns to the rule table with 24-bit subnet mask at this time. Due to the subnet mask of this rule table, QNS uses 255.255.255.0 to mask source IP from 192.168.88.32 to 192.168.88.0 and then calculates the hash value. Ultimately, it finds a SNAT rule on the basis of masked IP and wildcard port. Once finding a NAT rule, QNS stops searching for NAT rules.
\begin{figure}[ht!]
\centering
\includegraphics[width = .48\textwidth]{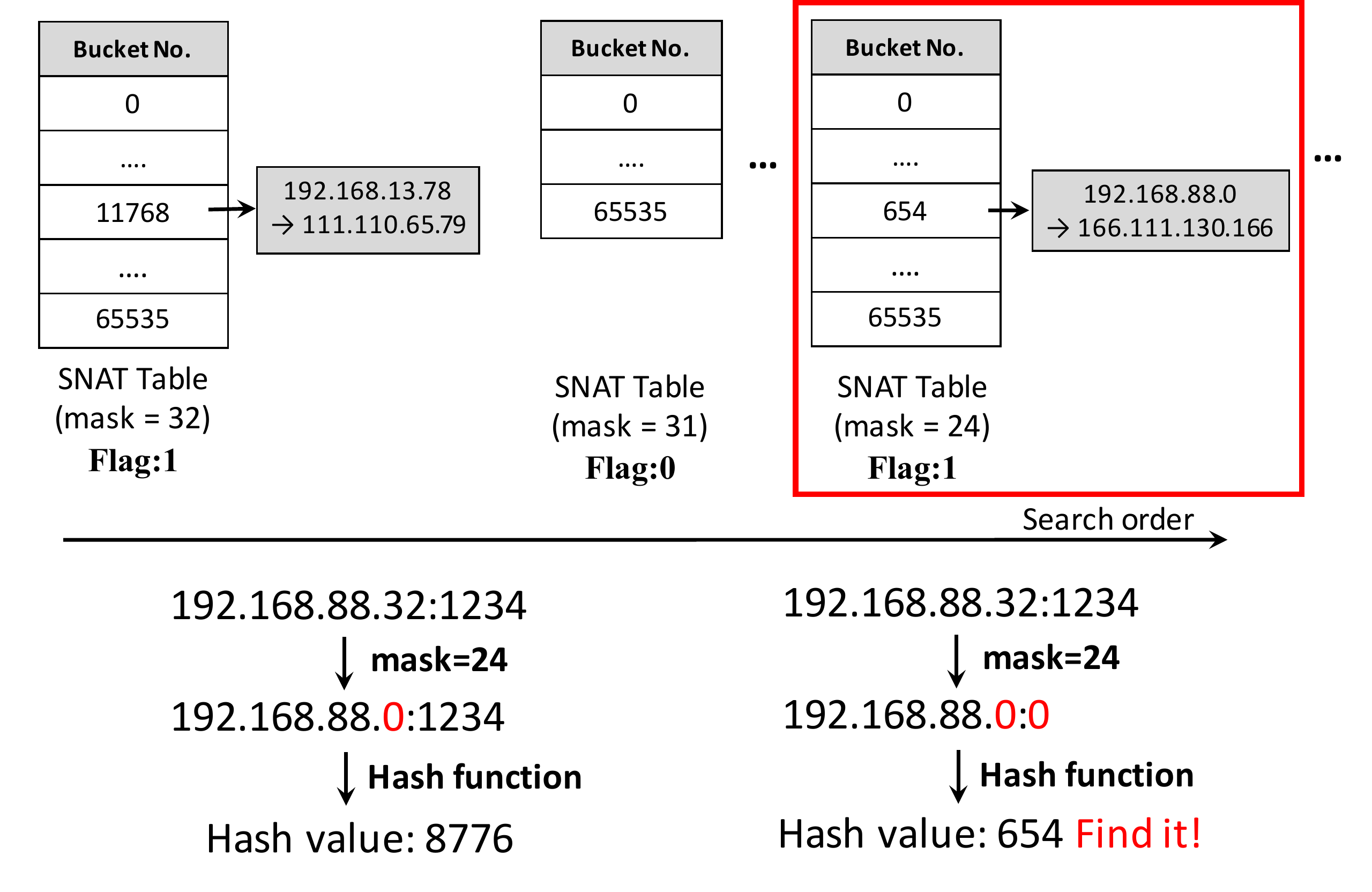}
\caption{Search for SNAT rule (step 3)}
\label{RSG}
\end{figure}

\subsection{Lock-free Sharing Among CPU cores}
Multicore processors have been pursued to improve overall processing performance. In Quick NAT system, connection record table is shared among CPU cores and built with lock-free hash table to eliminate the overhead of locks and make it more efficient and scalable to share connection records on multicore server.

\subsection{Polling and Zero-copy Packet Delivery}
We take advantage of Intel's DPDK to poll for data from NIC to eliminate overheads existing in interrupt-driven packet processing. In addition, Quick NAT manipulates the packet through pointers without copy operations in the process of NAT, increasing the throughput of packet processing.

\section{Evaluation}
In our experiment, we use three Intel Core CPU i7-5930k @ 3.5GHz (6 cores) servers -- one for the Quick NAT system under test and another two acting as traffic generator and receiver -- each of which has an Intel 82599ES 10G Dual Port NICs and 16GB memory. Ubuntu 16.04 (kernel 4.4.0) and DPDK-16.11 are used. We use Wind River's DPDK-pktgen \cite{pktgen} to generate traffic in line rate. 

We add a number of rules and then look for different NAT rules for 100 times to calculate the mean searching time of QNS algorithm in Quick NAT system and linear search algorithm that Netfilter uses to search for NAT rules. We change the number of rules  up to 10k and do this experiment for many times.

\begin{table}[!htbp]
\centering
\caption{The time of rule lookup (ns)}
\begin{tabular}{c|ccccc}
\hline
Rule Number & 100 & 1000 & 3000 & 5000 &10000\\
\hline
QNS Algorithm & 43 & 45& 42& 45 & 45 \\
Linear Search & 420 & 3256 & 9373 & 15078 & 30120\\
\hline
\end{tabular}
\end{table}

We can learn from the result in table I that it takes about 43 ns for QNS algorithm to search for rules and that the number of rules makes no difference on the performance of QNS algorithm because QNS is based on hash search with the complexity of O(1). On the contrary, it is a time consuming process to use linear search to look for rules in Netfilter, especially when the number of rules is large.

\bibliographystyle{unsrt}
\bibliography{./Quick_NAT_bibli}

\end{document}